\documentclass{aa}
\usepackage{graphics}
\def\hi{H\,{\sc i} }

\def\kmss{km~s$^{-1}$ }
\def\kms{km~s$^{-1}$}

\def\msol{${\rm M}_{\sun}$}
\def\smppc2{${\rm M}_{\sun} {\rm pc}^{-2}$}
\def\rcore{R_{\rm core}}
\def\kprime{K^{\prime}}

\begin{document}
\title{Dark and luminous matter in the NGC 3992 group of galaxies}
\subtitle{I. The large barred spiral NGC 3992}
\author{Roelof Bottema\inst{1} \and Marc A.W. Verheijen\inst{2}}
\offprints{R. Bottema}
\institute{Kapteyn Astronomical Institute, P.O. Box 800, 
NL-9700 AV Groningen, The Netherlands, e-mail:robot@astro.rug.nl
\and University of Wisconsin, Department of Astronomy, 
475 North Charter Street, Madison, WI 53706, U.S.A., 
e-mail:verheyen@astro.wisc.edu}
\date{Received date1; accepted date2}
\abstract{
Detailed neutral hydrogen observations have been obtained of
the large barred spiral galaxy NGC 3992 and its three small companion
galaxies, UGC 6923, UGC 6940, and UGC 6969. 
For the main galaxy, the \hi distribution is regular with a
low level radial extension outside the stellar disc. However, at exactly
the region of the bar, there is a pronounced central \hi hole in the
gas distribution. Likely gas has been transported inwards by the bar
and because of the emptyness of the hole no large accretion events
can have happened in recent galactic times. The gas kinematics is
very regular and it is demonstrated that the influence of the bar
potential on the velocity field is negligible. 
A precise and extended rotation curve
has been derived showing some distinct features which can be
explained by the non-exponential radial light distribution of NGC 3992.
The decomposition of the rotation curve gives a slight preference
for a sub maximal disc, though a range of disc contributions, up to
a maximum disc situation fits nearly equally well. For such a
maximum disc contribution, which might be expected in order
to generate and maintain the bar, the required mass-to-light ratio is large
but not exceptional.
\keywords{galaxies: individual: NGC 3992 --
ISM: kinematics and dynamics --
galaxies: kinematics and dynamics --
galaxies: spiral}
}
\maketitle
\section{Introduction}

Rotation curves derived from neutral hydrogen observations at
the outer regions of spiral galaxies unambiguously show that
substantial amounts of dark matter are required (Bosma 1978;
Begeman 1987, 1989). Any physically reasonable distribution
of this dark matter necessitates the presence of at least some of
that in the inner optical disc region, contributing in some degree
to the total rotation in that region. Unfortunately, from the
observed rotation curve and light distribution one cannot a priori
determine the ratio of dark to luminous matter (van Albada et al. 1985).
There are arguments, mainly theoretical, that the contribution of
the disc has to be maximized, leading to the so called maximum
disc hypothesis (van Albada \& Sancisi 1986; Salucci et al. 1991;
Sellwood \& Moore 1999). On the other hand, observations of disc
stellar velocity dispersions (Bottema 1993, 1997) lead to the conclusion
that the disc contributes, on average, 63\% to the total rotation
at the position where the disc has its maximum rotation. This finding
is supported by a statistical analysis of rotation curve shapes in
relation to the compactness of discs (Courteau \& Rix 1999).
Every detailed rotation curve of any galaxy may give clues as to
the ratio of dark to luminous matter in a galaxy. It has been
argued that the observed correlation of features in the rotation curve with
features in the photometry excludes a sub maximum disc case
(van Albada \& Sancisi 1986). Since it was already 
known that NGC 3992 exhibits
such specific rotation curve features, this warranted more detailed
observations and a proper decomposition of the curve into the
contributions of the galactic constituents.

At least a substantial fraction of spiral galaxies has no bar.
Yet an isolated
cold stellar disc can never be stable (Ostriker \& Peebles 1973).
Various criteria for galaxies have
been put forward that should be obeyed in order to avoid a
bar instability. Toomre's (1964) $Q$ criterion for local stability
can be applied for a global situation assuming a minimum $Q$ 
value for a stellar disc (Sellwood \& Carlberg 1984).
Already in 1973 Ostriker \& Peebles showed by numerical experiments
that a substantial spherical dark halo can stabilize a disc. Their
criterion states that ${\langle} E_{\rm kin} \rangle / E_{\rm pot} < 0.14$,
or the ratio of the average kinetic energy of a disc to its potential
energy should be less than 0.14 in order to be stable.
As an alternative to a dark halo, a disc can be stabilized
by a substantial bulge (Sellwood \& Evans 2001). In addition
to adding potential energy the bulge also makes the rotation
curve flat in the inner regions, creating an ILR which may inhibit
the bar formation mechanism (Toomre 1981). Calculations
for a specific set of galaxies by Efstathiou et al. (1982) showed
that for

\begin{equation}
\varepsilon = \frac{v_{\rm max}}{\sqrt{G M_{\rm disc} /h}} \la 1.1,
\end{equation}
the disc was unstable to bar formation. This criterion has recently
been reviewed and reanalysed by Syer et al. (1999) for a 3D situation.
For discs with a ratio of scalelength ($h$) to sech-squared thickness
$z_0$ of five and $Q$ = 1.2 at $R = 2.4 h$ they found that the numerical
value of 1.1 in Eq. (1) has to be decreased to 0.7. Though this criterion
may have its merits, it is difficult to imagine that disc stability does
not depend on its stellar velocity dispersion. Hence it may be dangerous
to apply it to a randomly observed galaxy.

As one can see, a robust, well established stability criterion
does not (yet) exist. Nevertheless, from the criteria just mentioned
some general relations can be extracted.
A galaxy becomes more unstable when its disc mass to dark matter ratio,
or its disc to bulge mass ratio is larger.
It is also more unstable when its stellar velocity dispersion
is lower. In the extreme one then has a cold, thin, maximum disc
situation, which is certainly unstable. 
Even a disc with $h/z_0$ of five, a representative
value for spiral galaxies (van der Kruit \& Searle 1982), at the maximum
disc limit forms a large bar (Bottema \& Gerritsen 1997).
A fair fraction of galaxies has a bar. At least for the long lived
bars it might then be logical to assume  
that barred galaxies are close to maximum disc while non-barred
galaxies are sub maximum, though there is no observational
evidence for this assumption. Yet the situation must be more complicated; the
stellar velocity dispersion is a factor of importance as is the gas
content and formation history of a galaxy.

From basic principles it can be shown that every bar has to
end within the corotation radius (Teuben \& Sanders 1985). This implies
that the angular pattern speed of the bar (${\Omega}_p$) always has
to be smaller than the angular speed at corotation (${\Omega}_{\rm cr}$).
A bar is defined as fast when its pattern speed approaches the
speed at corotation; it is slow when ${\Omega}_p \ll {\Omega}_{\rm cr}$.
Theoretical arguments related to the observed morphology of bars
indirectly lead to the conclusion that bars must rotate fast
(Sanders \& Tubbs 1980; Athanassoula 1992). Recently, by applying the
Tremaine \& Weinberg (1984) method it has been demonstrated directly
by observations of two galaxies (Merrifield \& Kuijken 1995; Gerssen
et al. 1999) that the bar is indeed a fast rotator.

Observations of neutral hydrogen gas show that in general there
is a depression in the \hi surface density at the position of the bar.
This can be explained because gas in a barred potential will
experience strong shocks at the leading side of the bar. As a result
gas will be transported inwards (Athanassoula 1992) and the bar region
gets depleted of gas.
The existence of a severe \hi hole then means that for a reasonable
amount of time the central region is not disturbed by gas accretion. 
In addition the bar must be rather long lived in order to 
transport all the gas to the centre. 

A barred disc does not exist on its own but is embedded
in a dark halo which should respond to the barred potential in
some way. Starting with a fast rotating bar in a nonrotating isotropic
dark matter halo Debattista \& Sellwood (1998) showed that the
bar is quickly slowed down by dynamical friction. At least when the
dark halo contributes significantly to the mass in the inner region;
for a maximum disc situation this slowing down mechanism has almost
disappeared. The initial setup of Debattista \& Sellwood is, of course,
rather specific. In reality a galaxy will gradually built up and
dark matter will acquire rotation by a number of mechanisms
(Tremaine \& Ostriker 1999). If the dark halo is co-rotating with
the bar the dynamical friction process has disappeared. Still it has
to be kept in mind that a fast rotating bar cannot exist in a
substantial non-rotating dark halo.
At this stage let us summarize a number of properties and processes
that are important for the existence of a barred structure.
Preventing a bar or destroying it can be done by:

\begin{itemize}
\item
A large non-corotating spherical dark matter (DM) contribution.
\item
A substantial bulge.
\item
Large amounts of gas will destroy any triaxiality (Shlosman \&
Noguchi 1993).
\item
A high stellar velocity dispersion. 
\end{itemize}
Making or allowing a bar can be done by:
\begin{itemize}
\item
A limited amount of spherical dark or luminous matter in the inner regions.
\item
Rotating dark matter.
\item
Low stellar velocity dispersions or alternatively a thin disc.
\item
Small amounts of gas (see also Noguchi 1996).
\end{itemize}

There are a number of neutral hydrogen observations of barred galaxies
preferentially of late types. Without trying to be complete we mention
NGC 5383 (Sancisi et al. 1979), NGC 4731 (Gottesman et al. 1984),
NGC 3359 (Ball 1986), NGC 1365 (Ondrechen \& van der Hulst 1989),
NGC 1097 (Ondrechen et al. 1989), NGC 1300 (England 1989), and
NGC 1073 (England et al. 1990). Observations of NGC 3992 were carried
out by Gottesman et al. (1984, hereafter G84) with the VLA. 
Their FWHM resolution amounted
to 23\arcsec\ in the spatial direction and 41 \kmss in
the velocity direction. They noted a drop in the rotational velocities
near the end of the disc light and this drop was ascribed to the
effect of a truncation of the disc mass. It is claimed that at least
some gas was detected in the bar region with a certain radial velocity,
although this might be an artifact of the way the data have been
interpreted. The velocity field and fit to that resulted in a
systemic velocity of 1046 \kms, a position angle of 248\degr, and
an inclination of 53\degr. A rotation curve was determined out to a
radius of 5\arcmin.

In a later study Hunter et al. (1988) made an additional analysis of
these observations. Their aim was to explain the observed spiral arm
pattern as a result of gas moving in the barlike potential. A model for
NGC 3992 was used employing a rigid potential consisting of a dark halo,
a disc, a bar, and an oval distortion beyond the perimeter of the bar. 
It appeared that this model was
not able to explain the observed tightly wound, star-formation arms.
Instead a strong two armed spiral structure was generated with large
pitch angle. Hunter et al. concluded that: ``either the model is incomplete
or, other, nondynamical processes cause the spiral arm pattern''.

NGC 3992 was also observed in \hi by Verheijen (1997) 
and by Verheijen \& Sancisi (2001)
as part of a study of the Ursa Major cluster. 
In fact, NGC 3992 is one of the most
massive members of this small and non-concentrated cluster. Because of
the limited integration time Verheijen could only study this galaxy
at a resolution of one arcminute. Consequently the derived rotation curve
was sparsely sampled.
Yet, these observations showed that the rotation curve exhibits
some specific features and that neutral hydrogen gas was present
rather far beyond the optical edge. When comparing the derived
M/L ratios of the galaxies in the Ursa Major cluster, the M/L ratio
of NGC 3992 stands out, in the sense that it is a factor
of two larger than the average value of the cluster galaxies.
These facts motivated a much longer observation aiming to 
shed more light on several matters. The main aim of this project
was to derive a high quality and extended rotation curve
for a barred galaxy. Of this rotation curve a decomposition
can be made and the result can be compared with that of other
galaxies. For example, is there evidence for a specific luminous
to dark mass ratio and does that differ from the ratio for
non-barred galaxies? Can the large M/L ratio be confirmed
and if so can it be explained? More generally, the number of well
determined rotation curves is still limited and extension
of the sample allows a better (statistical) analysis of galaxy
parameters. 

For convenience a high quality photograph of NGC 3992
is presented in Fig.~1 and a listing of the main parameters of the four
galaxies is given in Table~1.
The distance to the UMa cluster as a whole and to NGC 3992
in particular has not yet been determined precisely.
Sakai et al. (2000) give a distance to the UMa cluster of 20.7 $\pm$ 3.2 Mpc
following from a Tully-Fisher analysis using the cepheid distances
to local galaxies. On the other hand, for a similar analysis,
Tully \& Pierce (2000) derive a distance of
18.6 Mpc, probably with the same error as that of Sakai et al.
In a recent re-evaluation of the HST distance scale project
(Freedman et al. 2001) the distances to the local calibrator
galaxies have decreased by $\sim$ 5\% and consequently the distances
to UMa of 20.7 and 18.6 Mpc should also be decreased by that amount. 
As for now, a distance of 18.6 Mpc seems reasonable and has been
adopted in the present paper. This distance differs, however,
from the 15.5 Mpc used in earlier studies of the UMa cluster
by Tully et al. (1996) and by Verheijen (1997).

\begin{figure}
\resizebox{\hsize}{!}{\includegraphics{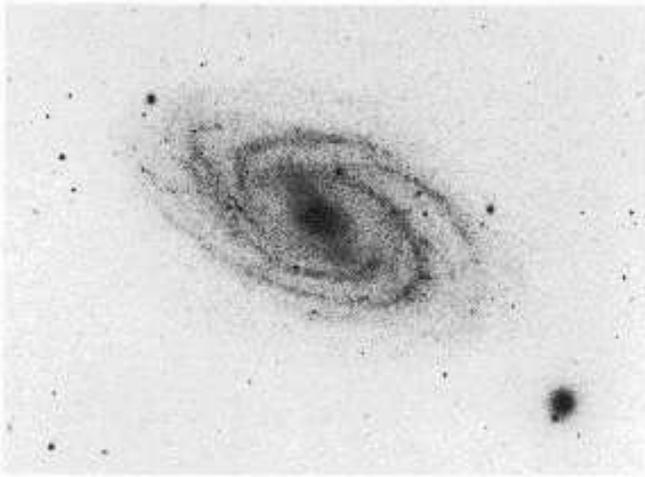}}
\caption[]{Optical image of NGC 3992 reproduced from the
Carnegie atlas of galaxies (Sandage \& Bedke 1994).
The photograph was taken with a blue sensitive emulsion.
Total size on the sky is {10}\farcm{7} $\times$ {7}\farcm{7},
North at top, East on left.}
\end{figure}

\section{Observations and data handling}

NGC 3992 and its companions were observed with the 
Westerbork Synthesis Radio Telescope in the period May 1997 
to September 1997. The observations lasted for approximately
48 hours divided up into 4 $\times$ 12 hour periods with different
antenna spacings. Due to maintenance, on average, two of the
fourteen telescopes, one movable and one fixed, were not available, 
which reduces the number
of interferometers from 40 to approximately 27. 
The digital line backend was configured for 64 frequency channels,
evenly spaced over a 5 MHz total bandwidth. After Hanning smoothing
this resulted in a velocity resolution equal to twice
the channel spacing, or 33 \kms, FWHM. A full listing of the
observing parameters is given in Table~2. The NFRA reduction
package NEWSTAR was used to do the calibration and fourier transform
to a 512 $\times$ 512 grid with a pixel size of 
{4}\farcs{32} $\times$ {5}\farcs{39} $(\alpha \times \delta)$.
Subsequent data reduction was performed with the GIPSY
(Groningen Image Processing SYstem) package.
 
Line emission of the main galaxy and the three companions was 
detected in 31 channels. Continuum emission only, was in principle
available in 15 channels at the low velocity side of the emission
and in 10 channels at the high velocity side. The continuum subtraction
was complicated by a strong continuum source at a distance of 
{34}\farcm{0} from NGC 3992, at 
position 11$^{\rm h}56^{\rm m} 39^{\rm s}$: 54\degr 9\arcmin
{48}\farcs{8} (RA:dec:1950) with a flux of 1.7 Jy. 
First, this continuum source
was cleaned away by subtracting 50 components with a gain of 0.5
found in a small region around this continuum source. This removed
the worst of the grating rings. 
The rest of the continuum was subtracted by fitting a linear
relation to the line free channels at both velocity sides and 
subtracting the appropriate amount from each channel in between. 
Next the data were cleaned. Clean components were subtracted until a 
level of 65\% of the noise level. Data were restored by convolving
the clean components with a Gaussian beam of 14\arcsec $\times$
18\arcsec (FWHM, $\alpha \times \delta$) and adding the residuals. 
In this way a cleaned collection of line channels, or data cube,
at full resolution was constructed, ready for further analysis.
To give an impression of the \hi intensity levels, \hi extensions
and of the position of the companions with respect to the main
galaxy, already at this stage a total \hi image at full resolution
of the whole field is presented in Fig. 2. A detailed description of 
how this image was constructed can be found in the next section. 

\begin{figure}
\resizebox{\hsize}{!}{\includegraphics{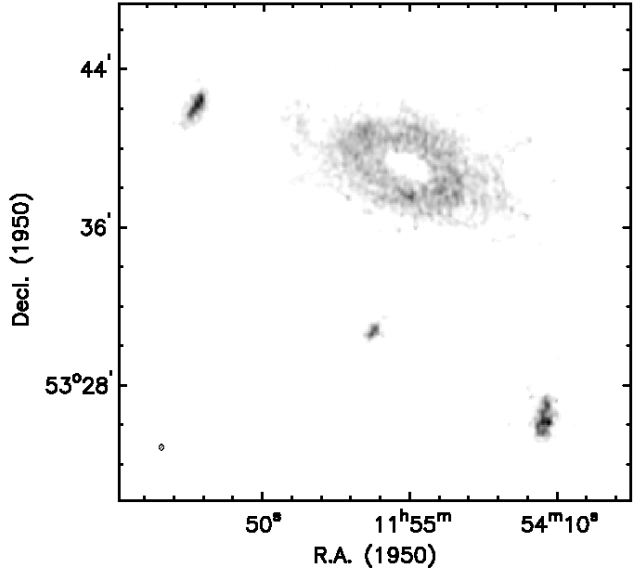}}
\caption[]{Greyscale image showing the full resolution
total \hi map of NGC 3992 and its surroundings.
From top left to bottom right the three companions,
UGC 6969, 6940, and 6923 are clearly visible;
their \hi column densities are larger than that of
the main galaxy. Note the central \hi hole of
NGC 3992, at the region of the bar.
The beam is indicated in the lower left, the greyscale
is linear from 0.2~10$^{20}$ to 34.8~10$^{20}$ H-atoms cm$^{-2}$.}
\end{figure}

The whole field was searched in detail for \hi emission other
than that from the already known sources. Nothing was found;
no unresolved sources above the noise level and no extended sources,
nor in the full resolution field nor after smoothing to lower 
resolution. Considering the noise statistics of the channel
maps, the threshold for detecting unresolved sources was put
at five times the rms noise level corresponding to \hi cloud masses
of 3.5~10$^6$ \msol.
As can be seen in Fig. 2 the three companions clearly
stand out in \hi and display a higher column density than the main galaxy. 
Thus if there is \hi in the field it is associated with a stellar
component, there are no free floating \hi clouds around, at least
not above the present detection limit. It can also be seen that the main
galaxy has a faint gas extension around the stellar disc, this contrary to
the companions, where the gas ends suddenly at the optical edge. 
A possible explanation for this is stripping of the gas from
the companions when these have passed by, or interacted with
NGC 3992.

\section{The \hi distribution}

\begin{figure*}
\resizebox{\hsize}{!}{\includegraphics{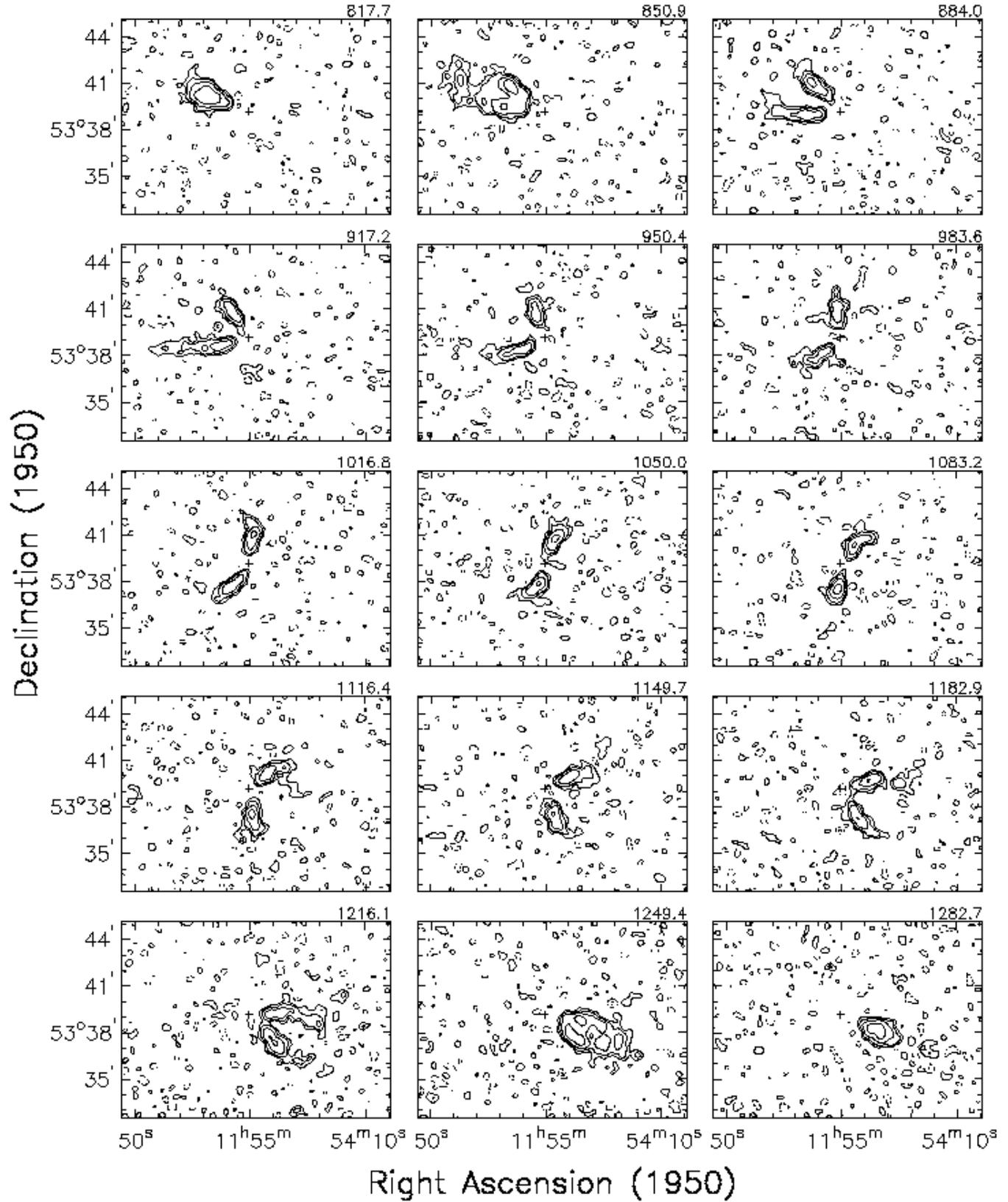}}
\caption[]{Mosaic of a number of uncorrelated cleaned channel maps
of NGC 3992 at a resolution of 30\arcsec $\times$ 30\arcsec.
The velocities of the channels are given at the top right and 
the cross indicates the position of the dynamic centre.
Contours are at levels of $-4{\sigma}, -2{\sigma},$ (dashed) 
$2{\sigma}, 4{\sigma}, 8{\sigma}, 16{\sigma},$ and $32{\sigma},$
where $\sigma$ is the rms noise level of the channels at 0.55~K
implying a level of 0.166~10$^{20}$ H-atoms cm$^{-2}$.}
\end{figure*}

A collection of uncorrelated channel maps smoothed to a resolution
of 30\arcsec $\times$ 30\arcsec\ is displayed in Fig. 3. The smoothed 
channels are displayed because only then the gas in the outer regions 
becomes visible. When inspecting Fig. 3 one can note the regularity of
the system. At first glance it looks like the rotation in the outer regions
is lower than in the luminous disc region. This is because emission in the 
disc region already shows up at velocities further away from the systemic
velocity than the emission of the outer regions.  

A total \hi map has been constructed at full resolution and at
30\arcsec $\times$ 30\arcsec\ resolution, both by means of the
conditional transfer method. Details of this method will now be described
for both resolutions. First the full resolution map. 
At positions in the 30\arcsec $\times$ 30\arcsec\ map where the
intensity level was higher than three times the $1\sigma$ noise 
level in that map,
the data in the full resolution map were retained. Data not meeting
this criterion were set to zero. In addition, all remaining positive 
noise patches in the full resolution maps were inspected, whether
above the five sigma level or whether extended. Unresolved patches
below this five sigma level were deleted. The remaining signal in each
channel was summed to give the flux density as a function of
velocity, or the full resolution line profile in Fig. 4. This is
the typical double horned profile as observed for normal spiral galaxies. 
Summing the emission in the data cube along the velocity direction
gives the total \hi map, shown in Fig. 5, top panel and as a greyscale
already given in Fig. 2. 

\begin{figure}
\resizebox{\hsize}{!}{\includegraphics{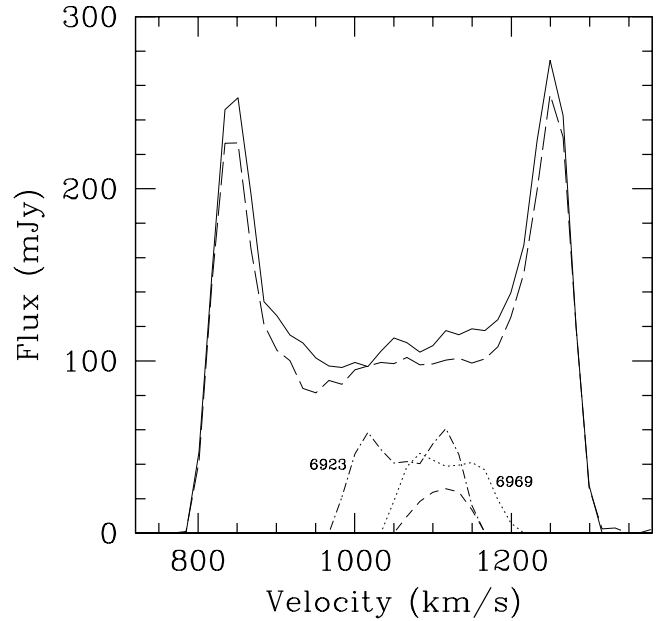}}
\caption[]{The \hi profiles of the four galaxies. The two large
profiles are for NGC 3992; at full resolution (dashed) and
at 30\arcsec $\times$ 30\arcsec\ resolution (full drawn). The
smaller profiles are for the companions UGC 6923, 6940 (dashed), and
UGC 6969. When placed at a distance of 18.6 Mpc the corresponding
total \hi masses for NGC 3992 (smooth), UGC 6923, 6940, and 6969
are 5.9, 0.64, 0.16, and 0.44~10$^9$~$M_{\sun}$ respectively.}
\end{figure}

\begin{figure}
\resizebox{\hsize}{!}{\includegraphics{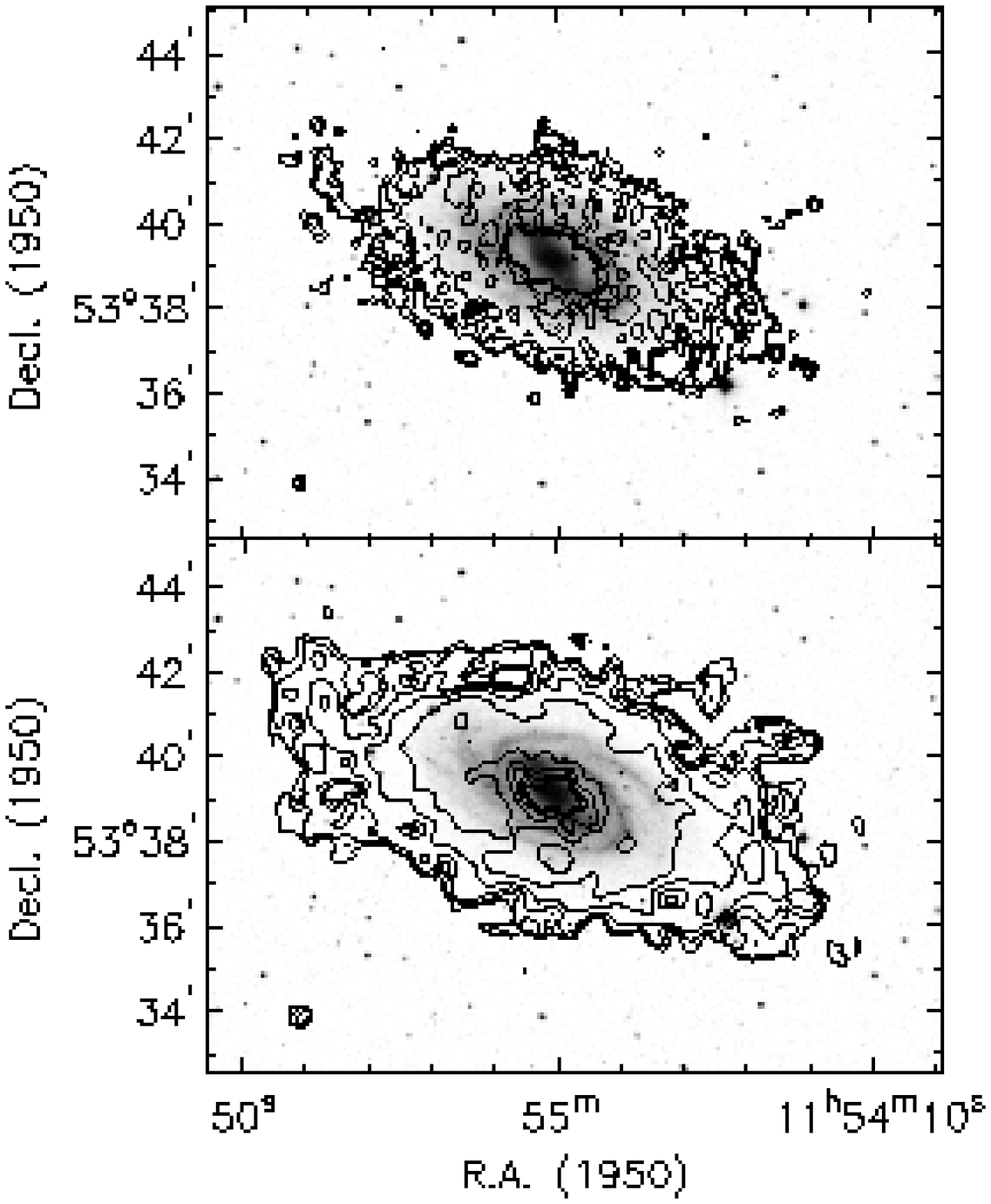}}
\caption[]{Total \hi map of NGC 3992 superposed on
the optical image. {\it Top:} 
At full resolution. Contour levels successively increase
by a factor two from 1.39~10$^{20}$ to 22.16~10$^{20}$ H-atoms cm$^{-2}$.
{\it Bottom:} At a resolution of 30\arcsec $\times$ 30\arcsec. 
Contour levels increase by a factor of two from 0.39~10$^{20}$
to 12.4~10$^{20}$ H-atoms cm$^{-2}$.
The resolution is indicated in the lower left corner.}
\end{figure}

The \hi emission at 30\arcsec $\times$ 30\arcsec\ resolution was 
determined as follows: The data cube was smoothed to a resolution of
90\arcsec $\times$ 90\arcsec. Data in the 30\arcsec\ resolution map were 
retained there where in the 90\arcsec\ resolution map data were above
the three sigma level (= 0.045 $M_{\sun}$~pc$^{-2}$). 
As above, unresolved patches
below the five sigma level in the 30\arcsec\ maps were deleted. 
The line profile is displayed in Fig. 4, and as can be seen the smoothed
maps all contain slightly more emission than the full resolution maps. 
This is caused by some additional low level emission
that has surpassed the $3\sigma$ level at 90\arcsec\ resolution.
Adding up all the emission gives
a total \hi flux of 72.2 Jy~\kmss resulting in a total \hi mass of
5.9~10$^9$ \msol. For the full resolution these numbers are slightly 
lower at respectively an \hi flux of 62.9 Jy~\kmss and total \hi
mass of 5.1~10$^9$ \msol. Integration of the data cube along the
velocity direction gives the total \hi map at 30\arcsec\ resolution
displayed in Fig. 5, bottom panel. 

As can be seen in Fig. 5, at the lower resolution additional low level
emission shows up in the outer regions. The gas distribution is symmetric
with respect to the luminous structure. Large spiral arm density enhancements
can not been seen, nor density features which can be associated with
the bar. At the rim the distribution might be more patchy, especially
at the North West side where the \hi is more concentrated in clouds. 
One very obvious feature which can be recognized immediately, both
in Fig. 2 and in Fig. 5 is the central hole at exactly the region
of the bar. But how empty is this hole? 

Two qualitative tests have been done to determine this emptyness. 
First, of the full resolution data cube a position velocity (or x,v)
slice has been made through the data cube, along the major axis and
with a width of the size of the hole. Any emission would then show up
as a narrow filament at the position of the galaxy rotation curve. 
The x,v diagram was inspected visually and nothing could be detected. 
As a second test the fact was used that near the centre 
one expects the rotation curve to be steeply rising.
Therefore any emission in the hole
should be at nearly the same positions in the relevant channel maps and
to increase the signal-to-noise these channels can simply
be added. 
After the rotation curve was determined, 21 channels were selected 
for this test, 10 on either side of the channel with a
velocity of 1050 \kms. These channels were added and the result 
inspected. At the central position the level and noise characteristics
were equal to regions outside the galaxy, meaning that no emission
was detected. 

It is not straightforward to give a quantitative value for 
the upper limit of the surface density
in the region of the hole. Let's give it a try. 
One channel at full resolution has a 1$\sigma$ noise level of 
0.473 \smppc2. Adding $N$ channels which have been Hanning smoothed,
each having a noise of ${\sigma}_h$ gives a total noise of 
${{4}\over{\sqrt{6}}}\sqrt{N-{{3}\over{4}}} \ast {\sigma}_h$.
So adding ten channels all with the same noise gives a total noise
level of 2.35 \smppc2, which is for one position on the sky. 
One expects approximately that, if added, those ten channels would
fill half the hole, which can be covered by $\sim$ 16 beams. Then
a $1\sigma$ upper limit for the surface density in the hole is
found of approximately $2.35/\sqrt{16}$ = 0.6 \smppc2. 
There are other ways of reasoning
to estimate the upper limit, but all arrive at the same or at a larger number. 

To obtain the surface density as a function of radius,
the observed total \hi map has been averaged on elliptic annuli.
These annuli were given the same orientation as for the fit of a 
collection in tilted rings to the velocity field in order to derive
the rotation curve (see Sect. 6). For radii less than 200\arcsec\
the full resolution map was used with widths at the major axis 
of 10\arcsec\ and
for larger radii the smoothed map with widths of 20\arcsec. 
The average value of each ellipse was deprojected to face-on. 
The result for the two
sides separately and averaged is shown in Fig. 6. It can be
seen that the \hi emission 
for this galaxy is concentrated is a torus with a low level
extension to large radii.  
Note that the Holmberg radius is at 250\arcsec\ and at that 
radius the transition occurs from a high \hi density to the low level
extension.

\begin{figure}
\resizebox{\hsize}{!}{\includegraphics{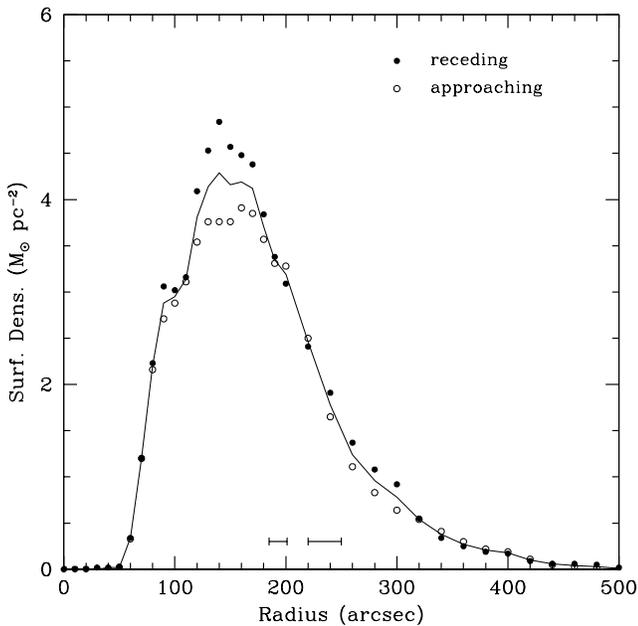}}
\caption[]{The deprojected \hi surface density as a function of radius.
The surface densities were obtained by averaging the total \hi map  
over elliptic annuli with the same orientations as
used for the rotation curve determination. For radii less than
200\arcsec\ full resolution data were used, elsewhere the smoothed data.
The \hi is distributed in a torus like structure with a shallow
extension at large radii. The deprojected bar radius is {72}\farcs{5}
so that the galaxy is devoid of \hi gas at the region of the bar.}
\end{figure}

\section{The continuum}

The continuum image has been made by averaging the channels free of line 
emission.
This image has been cleaned to below the noise level and smoothed to
a 30\arcsec $\times$ 30\arcsec\ resolution. Even then the emission level
at the position of the galaxy is low; the peak emission is at six
times the noise level of 0.18~K. The image is displayed in Fig. 7 in
greyscale and in contours superposed on the optical image. Despite the
low level of emission a few associations between continuum and the optical
picture can be made. There is a continuum enhancement at the position
of the bulge. Furthermore, continuum is associated with spiral arms
especially those at the South West side. A strong unresolved source 
with a flux of 1.5 mJy can
be seen in the South East on a spiral arm.
This source might originate from a star formation region though
an association with an optical counterpart is not obvious.
It is certainly too bright to be produced by a single supernova remnant. 
An alternative explanation is a background source. 
The total continuum flux of the galaxy amounts
to 43.2 mJy being quite normal for galaxies of the size of NGC 3992. 

\begin{figure}
\resizebox{\hsize}{!}{\includegraphics{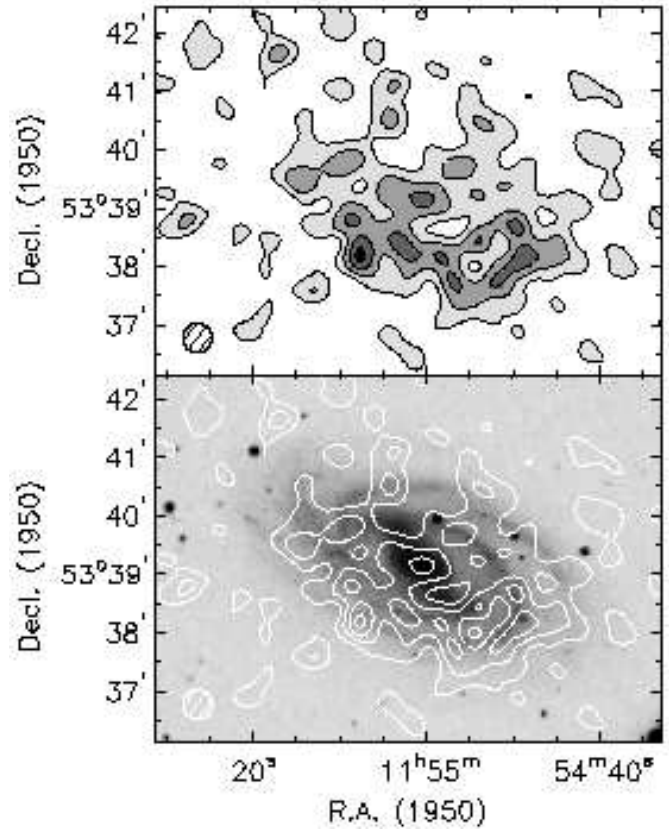}}
\caption[]{Cleaned continuum map of NGC 3992 at a resolution of
30\arcsec $\times$ 30\arcsec. The continuum emission is at a low level
and consequently has a noisy appearance.
{\it Top:} A contour and greyscale map. Contour levels are
at 1.5, 3, 4.5, and 6 times the rms noise level of 0.18~K.
{\it Bottom:} As a contour map superposed on the optical
image. The total continuum flux of 
NGC 3992 amounts to 43.2 mJy.}
\end{figure}

\section{Construction of the velocity field}

\begin{figure*}
\resizebox{\hsize}{!}{\includegraphics{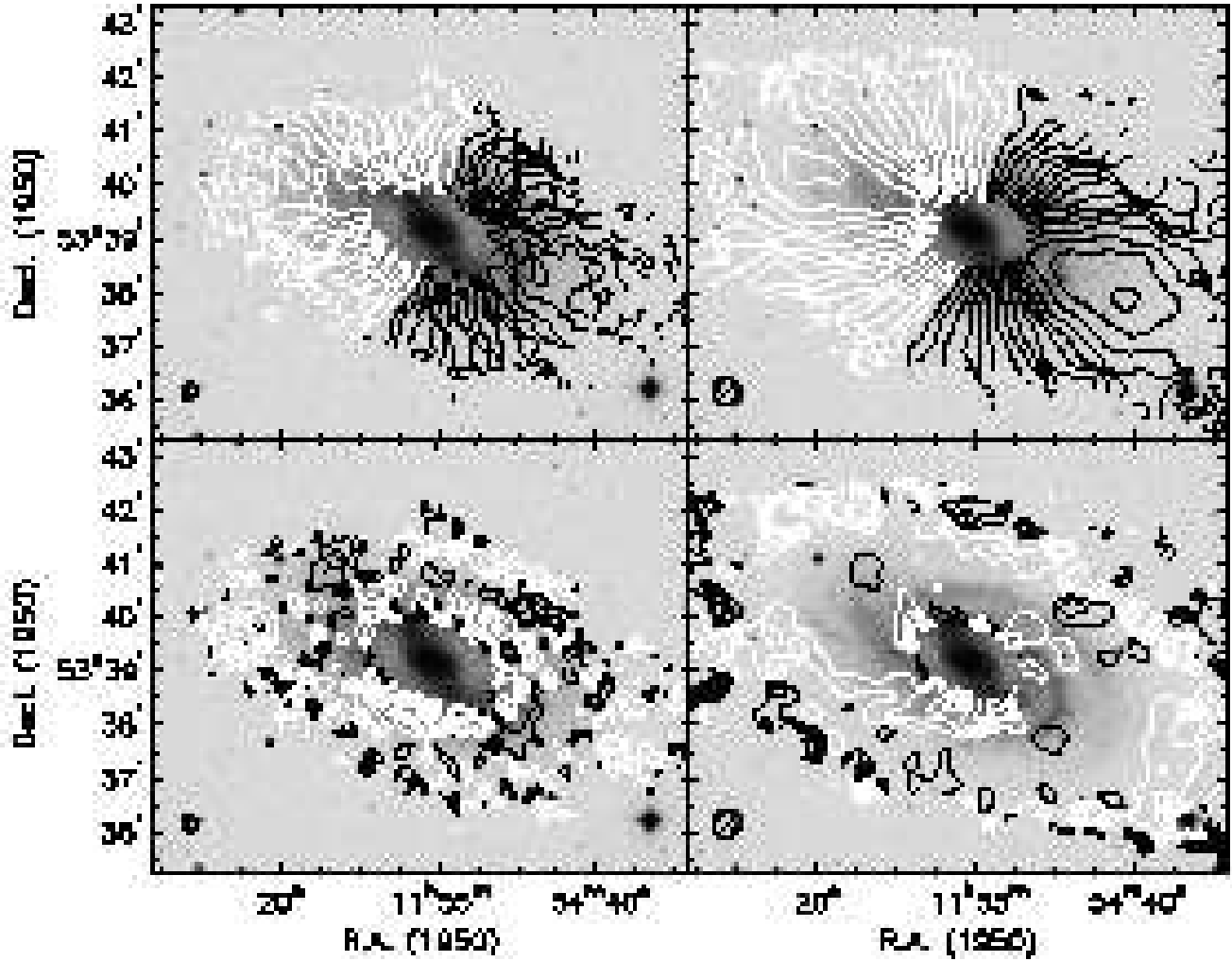}}
\caption[]{Optical image of NGC 3992 with superposed the
velocity field at full resolution (top left), the velocity field
at 30\arcsec $\times$ 30\arcsec\ resolution (top right), the
residual velocity field at full resolution (bottom left),
and the residual velocity field at 30\arcsec $\times$ 30\arcsec\
resolution (bottom right).
For the velocity fields the systemic velocity of 1049 \kmss is
at the first black contour next to the white contours.
Contours differ by 15 \kmss and increase from left to right.
For the residual maps contours differ by 5 \kms,
white is negative, black positive velocities. The residual velocity
field is obtained by subtracting a model velocity field determined by
a tilted ring fit (see Fig. 9) from the
observed field.
One may notice streaming motions along the spiral arms,
especially on the North-West in the full resolution velocity field.
Only near the rim of the central hole there are some
systematic residuals that may be attributed to the bar potential.}
\end{figure*}

At first, several line profiles distributed over the galaxy
have been inspected by eye. It appears that almost all profiles
are symmetric. In addition there are no regions in the galaxy where
there is a systematic skewness of the profiles in any velocity direction. 
Hence there is no need to make a fit by a model profile other than
that of a Gaussian. 

Two velocity fields were constructed, one at full resolution and one
at 30\arcsec $\times$ 30\arcsec\ resolution. For both fields the method of
construction was equal and will be described presently. The Gaussian fitting
procedure needs decent initial estimates for the profiles. 
To that aim, a Gaussian
fit was made to the conditionally transferred channels. The resulting
parameters were then fed to the fitting procedure for the whole
data cube. In this way it is assured that only line profiles are found
there where it was judged already before where the \hi gas was situated. 
If the initial estimate had a dispersion of less than 5 \kmss
or a peak amplitude less than 1.5 the noise level, it was discarded. 

With these estimates a fit to the complete data cube was made. 
Again, results with dispersions less than 5 \kmss and amplitudes less
than 1.5 times the noise level were judged to be unphysical and were rejected. 
The velocity field was inspected by eye to check for continuity of the
data. It appeared necessary to remove only a small number, between 
10 and 20, deviating pixels for both velocity fields. These pixels either had
a strongly aberrant velocity or a velocity dispersion larger than 
65 \kmss and were nearly all situated at the low intensity edges. 

Of the resulting dispersions of the profiles, 70\% had a value between
5 and 25 \kms, 25\% between 25 and 45 \kms, and 5 \% above 45 \kms.
The instrumental FWHM velocity resolution of 33.3 \kmss equals an
instrumental dispersion $(1\sigma)$ resolution of 14.1 \kmss and
thus most of the profiles are not resolved in velocity
by the telescope. This is a reflection of the
inherently low gas velocity dispersion in galactic discs, between
6 and 12 \kmss (Kamphuis 1993; Dickey et al. 1990).

The velocity fields at both resolutions are displayed in 
Fig. 8, top half. In general the velocity field of NGC 3992 is 
astonishingly regular. Large warp or bar signatures are not present.
Some streaming motions can be observed along spiral arms, especially
on the North West side.
  
\section{The rotation curve}

The rotation curve is determined by fitting a tilted ring model
to the velocity field (Begeman, 1989). It is assumed that 
the \hi gas can be described
by a set of concentric rings. Each ring has a certain position, systemic
velocity, rotation velocity, and two orientation angles: inclination 
and position angle. The widths and separations 
of the rings were taken to be 10\arcsec\
for the full resolution data and 20\arcsec\ for the data with resolution
of 30\arcsec.  
When fitting the model velocity field to the observed field a weighting
factor has been taken proportional the the cosine of the angle 
measured from the major axis.

An iterative strategy was followed. First a guess has been made of
the rotation, inclinations, and position angles for all rings. With 
these parameters held fixed the central positions and the 
systemic velocities of the rings were determined. 
These parameters appeared to be nicely
constant as a  function of radius and were fixed at (RA, dec, $v_{\rm sys}$)
of (11$^{\rm h}$~55$^{\rm m}$~{0}\fs{59}, 53{\degr}~39{\arcmin}~{10}\farcs{9},
1049 $\pm$ 2 \kms).
The optical position of the bulge was measured
and lies within {0}\farcs{2} from the kinematic position while
both positions can be determined with an accuracy of approximately 2\arcsec.
Consequently the bulge is exactly at the kinematic centre
even though that was determined by an extrapolation inwards
because of the \hi hole. This proves the overall regularity
and symmetry of the velocity field. 
  
Further steps in the iteration are
illustrated in Fig. 9. 
As a first step the remaining parameters, PAs, inclinations,
and rotational velocities were left free and the
resulting position angles were considered. As
can be seen in Fig. 9, both for the full resolution and the smoothed data
the position angle slightly changes as a function of radius.
In addition, for the full resolution data there is a wiggle superposed
which is caused by the spiral arm streaming motions. In first instance
a constant position angle of 248\degr\ was assumed and rotation curve
determined. However, for that case the residual velocity field 
showed a large scale systematic pattern which could be attributed to 
a wrong position angle. Therefore it was decided to adopt a position angle
which is
slightly changing as a function of radius, as indicated by the dashed
line in figures 9b and 9d.   
Considering the errors on the data points this change seems indeed real.
Moreover, in this case the systematics in the residual velocity field
disappear.
As a second step the position angles were fixed and
the inclinations and rotational velocities were left as free parameters. 
The fitting procedure was rerun and resulting inclinations
were considered. For the full resolution data there appears to be
a slight increase in the inclination from 57 to 60\degr\ between radii
of 280 to 320\arcsec. However, for the smoothed data this increase 
is not present and this effect in the full resolution data probably
has to be ascribed to patchiness and small irregularities in the
velocity field at high resolution. 
Over all, the data are consistent with a constant inclination of
57\degr\ $\pm$ 1\degr\ indicated by the dashed line in figures 9a 
and 9c. This constant value has been adopted. 
As a third and final step in the iteration, 
having the PAs and inclinations fixed, the rotation velocity was fitted
of which the result
is displayed in Fig. 9, lower panel.

\begin{figure}
\resizebox{\hsize}{!}{\includegraphics{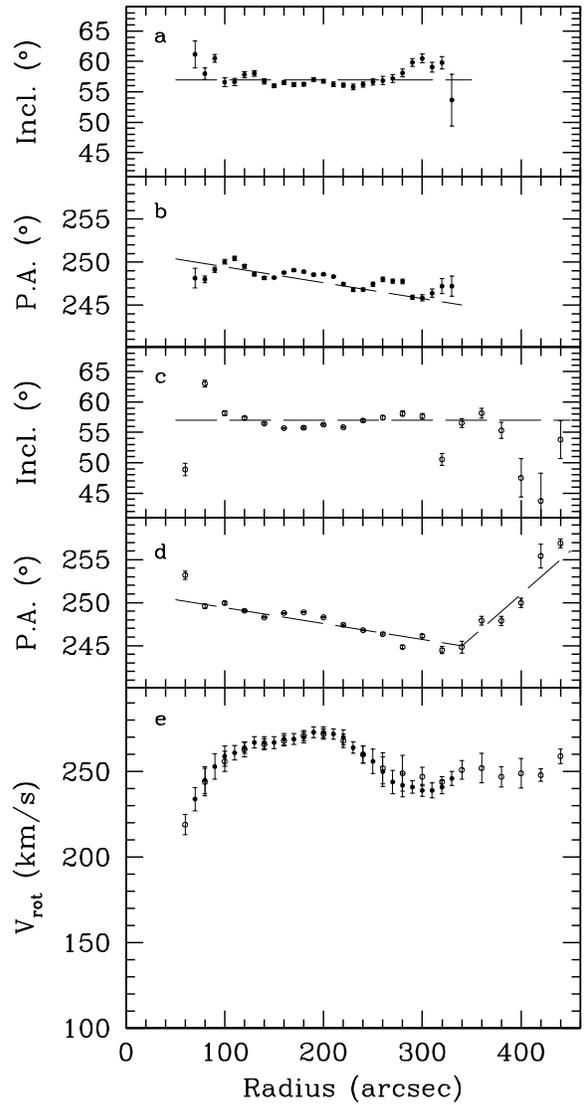}}
\caption[]{Determination of the rotation curve by a tilted ring fit;
filled circles for full resolution data
(a, b, and e), open circles for smoothed data (c, d, and e).
In first instance the orientation angles and rotation velocity
were all left as free parameters producing position angles
as a function of radius (b and d). These PAs were fixed
at the values indicated by the dashed line
and the fitting procedure was rerun producing the inclinations (a and c).
Fixing these at 57\degr\ finally gives the rotation curves (e).}
\end{figure}

At radii between 270 and 320\arcsec\ which is at the outermost radii where
the full resolution observations still give
rotational values, the full resolution
rotation velocities are lower by some 6 \kmss compared to the smoothed data.
A possible explanation is that at those positions the full resolution
only pics up the brightest emission regions at the spiral arms. Velocities
over there might deviate somewhat from the average velocities because of
streaming motions associated with the spiral arms. Also at these radii
the stellar disc ends, which might have some influence on the radial velocities
when changing from a situation of stars plus gas to pure gas arms.
Anyway, from 240\arcsec\ outwards the smoothed data have been adopted
as best representation of the rotation. Inwards from 240\arcsec\
until 70\arcsec, where the hole begins, the full resolution rotation
is determined with sufficient certainty. As a summary the rotation curve
data are given in Table 3.

The least squares fitting method gives errors, but these are only
formal errors, which are not always a good representation of
the true deviation from the data. To come up with a more realistic error
of the rotational velocity, the fitting procedure has been repeated
for the receding and approaching side of the galaxy separately.
Positions and orientation angles were kept fixed at the same values as
for the whole galaxy and rotation velocities were determined.
The difference in rotation of the two sides
gives a better representation of the true error.
The final error is then given by the quadratic sum of the formal fit error
plus half the difference between the two sides, plus the error generated by
an error of 1\degr\ in the inclination. These values are also given
in Table 3 and in Fig. 9. To illustrate the reliability and 
consistency of the method, in Fig. 10,
the rotation curve, converted to radial velocities is overplotted on
a position-velocity cut through the smoothed data along the major axis.
Note that the globally determined rotational values may deviate slightly
from the local kinematics given in x,v diagram.

The present rotational parameters have been compared with the ones
of Verheijen \& Sancisi (2001). They have produced a rotation curve
from the velocity field at a resolution of only 60\arcsec\ and
therefore they have a rather course sampling of the rotation curve.
Only five independent and reliable PA and inclination values could be
determined between 160 and 330\arcsec\ from which a constant value
of the orientation angles was concluded. 
The rotation curve deduced goes out to 400\arcsec\ and is qualitatively
the same as determined presently but shows a lot less details
and features.

A comparison was made with the observations and derived kinematics
in the paper of G84. They have observed NGC 3992 with the VLA
at a FWHM resolution of 23\arcsec\ spatially and of 41.4 \kmss
in velocity. Their rotation curve extends out to a radius of 300\arcsec\
and is similar in shape as the present curve. There is a 
difference, however; G84 determined an inclination of {53}\fdg{4}
which is smaller than our value of 57\degr. Consequently G84's
rotational velocities are a bit larger. It is claimed, though uncertain,
that there is a small amount of gas with associated radial velocities 
detected in the central hole. In our opinion this detection
is not real and is probably an artifact of the employed moment
method to construct the velocity field.

\begin{figure}
\resizebox{\hsize}{!}{\includegraphics{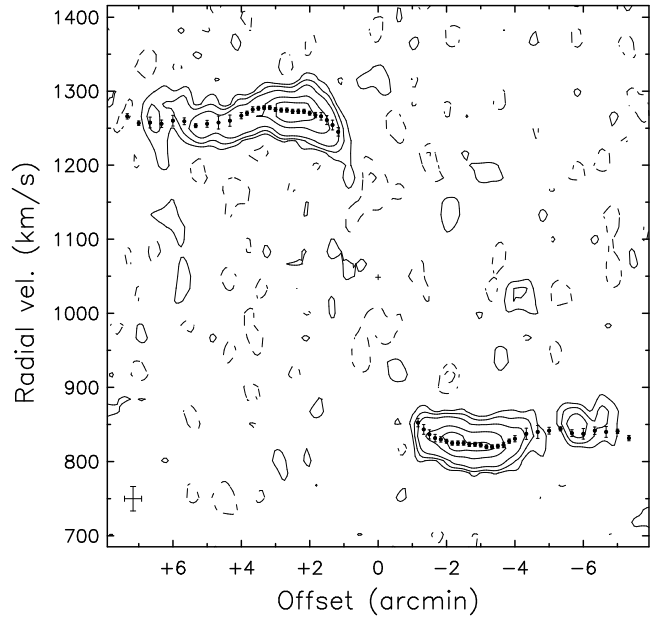}}
\caption[]{The rotation curve converted to radial velocities
overplotted on a position - velocity map along
the major axis. This illustrates 
the consistency of the tilted ring fitting method.
Contour levels are at -1.65 and -0.82~K (dashed),
and at 0.82, 1.65, 2.93, 6.59, and 9.89~K, and the resolution
(30\arcsec $\times$ 33.3 \kmss FWHM) is indicated in the
lower left corner.}
\end{figure}

A model velocity field has been created assuming the fitted rotation
curve parameters. This field was subtracted from the observed
velocity field producing the residual velocity field which is
depicted in Fig. 8.
As can be seen, in the full resolution case there are some residuals
associated with the streaming motions along the spiral arms. Other
systematic residuals are not present demonstrating that a proper
rotation curve fit has been made. The random deviations are
generally smaller than 10 \kmss except for a few patches bordering on
the central hole. Could these deviating patches have been generated
by the central bar? 

\section{The influence of the bar}

The dynamical influence of the bar has been determined by calculating
the potential field generated by the luminous distribution.
To that aim
the I-band image of NGC 3992 has been properly sky subtracted
and defects and stars were interpolated over. The image was
projected to face-on using a PA of 68\degr\ and inclination
of 57\degr\ following from the \hi kinematics.
As a bonus one can quite accurately 
measure the physical length of the bar in the face-on image which
amounts to 145{\arcsec}. Next the potential
image was calculated numerically and ellipses have been fitted to it. 
The resulting ellipticities 
$(\epsilon = 1-b/a)$
as a function of the major axis of
the ellipse are given in Fig. 11.
When inspecting Fig. 11 it is obvious that
the ellipticity of the bar potential is largest at a radius of 60\arcsec,
at smaller radii the bulge makes the potential rounder while at
larger radii the galactic disc takes over.

The bar ends at {72}\farcs{5} yet the extent of its
elliptic potential does not reach beyond approximately 90\arcsec.
As can be seen in Fig. 11, there is only a small overlap
between the elliptic bar potential and presence of \hi gas,
roughly between 70 and 90\arcsec, which is just at the position
of the deviating patches of the velocity field bordering
the central hole. It is therefore likely that these aberrations
are indeed generated by the bar. On the other hand, judging
the ellipticity of the potential, the bar will not generate
a distortion of the velocity field for radii larger than 90\arcsec.
Consequently the derived rotation curve is beyond dispute at those
radii. Even at the positions further in, the residuals of the
velocity field are so moderate that the rotational values are
still reasonably well determined.
%
\begin{figure}
\resizebox{\hsize}{!}{\includegraphics{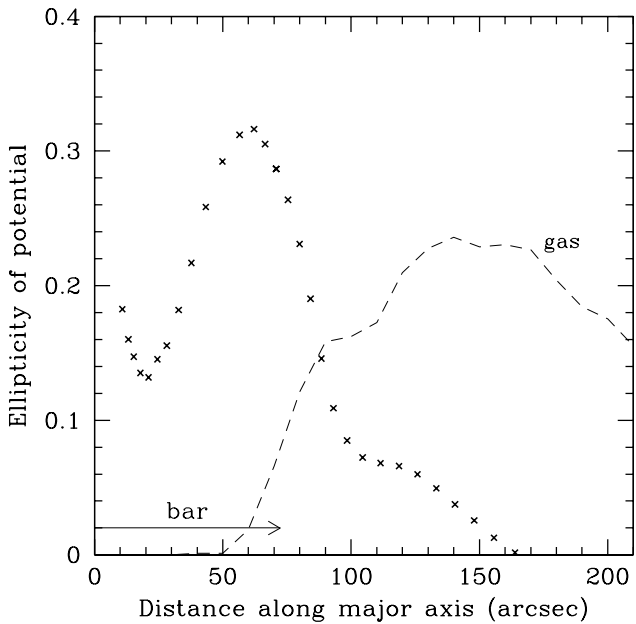}}
\caption[]{
The crosses give the
ellipticity of the potential as generated by the 
I-band image which was projected to face-on.
The influence of the bar barely reaches into the regions
where gas is present (dashed line). 
}
\end{figure}

\section{Decomposition of the rotation curve}

The observed rotation curve can be decomposed into the contributions
of the individual galactic mass constituents (van Albada et al. 1985;
Kent 1986, 1987; Begeman et al. 1991). As usual, for the luminous components
a constant M/L ratio is adopted. Then for each luminous component a mass model
can be constructed proportional to the light distribution which is
given by the photometry. The gravitational influence of the gas is
included and a dark halo is needed to explain the large rotation
in the outer parts of galaxies. The decomposition of the observed rotation
curve can then be achieved by a least squares fitting procedure where
the individual rotational contributions of disc, bulge, gas, and
dark halo are determined. 
Such a fit is, however, far from unique (van Albada et al. 1986);
in general a range of M/L ratios for the disc (and bulge) gives
an equally good fit to the data. Additional information or assumptions
are needed. One such assumption is the maximum disc hypothesis which
states that the contribution of the disc to the rotation should be
scaled up as high as possible and so assigning the maximum possible
M/L ratio to the disc. On the other hand, observations of the
stellar velocity dispersions of galactics discs suggest that the
maximum contribution of the disc to the rotation is, on average,
63\% at the position where the disc has its maximum velocity (Bottema
1993). As discussed in the introduction, 
there is other evidence that for normal and specifically for 
the low surface brightness
galaxies the maximum disc hypothesis cannot hold.
Below we will investigate the
possibilities for NGC 3992.

Photometry in the B, R, I, and $\kprime$ band is given by
Tully et al. (1996), of which the I and $\kprime$ data are reproduced
in Fig. 12. Since the $\kprime$ data do not extent very far out,
the I band data are used for the mass modelling. Note that in the
inner regions the I and $\kprime$ photometry are nearly identical
and hence there are no large population and dust gradients in the
inner regions and the I band gives a good representation of the
actual mass distribution. 
The total luminosity of NGC 3992 in the I band is identical to
within the errors with the total luminosity derived by H\'eraudau
\& Simien (1996) which gives confidence that the absolute calibration
of the surface brightness is correct.
The photometry has been cut off at a
radius of 250\arcsec\ but the calculated rotation curve does not
depend on the exact position of the cutoff as long as it is
at or beyond this 250\arcsec.

\begin{figure}
\resizebox{\hsize}{!}{\includegraphics{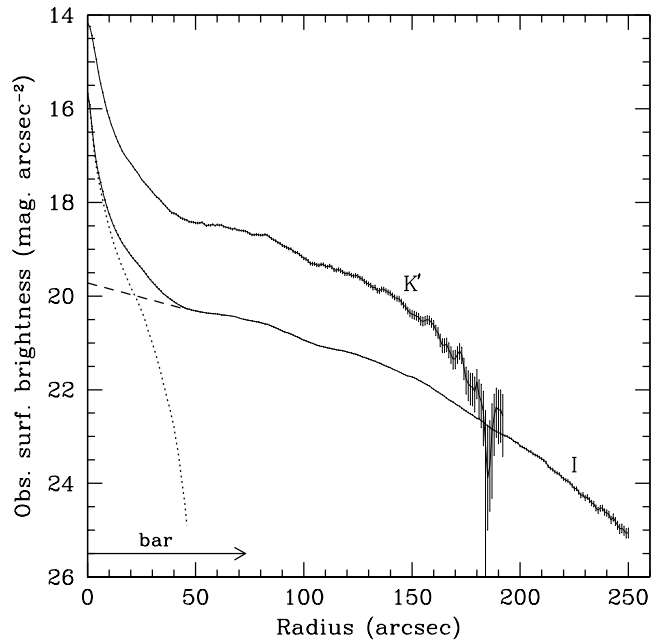}}
\caption[]{Observed photometric profiles in the I and $\kprime$ band
of NGC 3992 by Tully et al. (1996). The I profile
is used for the rotation curve decomposition. A
bulge/disc decomposition has been made
by assuming a disc profile (dashed line) which
has been subtracted from the observed profile to get
the bulge light (dotted line).}
\end{figure}

For the gaseous component a surface distribution equal to that of the \hi
is taken (see Fig.~6), multiplied with a factor 1.4 to account for Helium.
The gas disc is adopted to be infinitely thin.
For the dark halo a pseudo isothermal sphere generally provides
a satisfactory model (Carignan \& Freeman 1985) for which the
density distribution ${\rho}(R)$ is given by

\begin{equation}
{\rho}_{\rm h} = {\rho}_{\rm h}^0 \left[ 1 + \frac{R^2}{R^2_{\rm core}}
\right]^{-1},
\end{equation}
with a rotation law
\begin{equation}
v_{\rm h} = v_{\rm h}^{\rm max} \sqrt{
1 - \frac{R_{\rm core}}{R} \arctan \left( \frac{R}{R_{\rm core}}
\right) },
\end{equation}
where $R_{\rm core}$ is the core radius related to the maximum
rotation of the halo $v_{\rm h}^{\rm max}$ by
\begin{equation}
v_{\rm h}^{\rm max} = \sqrt{ 4\pi G {\rho}_{\rm h}^0
R^2_{\rm core} }.
\end{equation}

In this section two situations will be considered.
At first one where the luminous mass is all in a disc like distribution.
The disc is given a locally isothermal sech-squared vertical distribution
(van der Kruit \& Searle 1981) with a $z_0$ scale height parameter of
700 pc. Secondly a decomposition of the photometric profile is made
in a bulge and a disc. This is done by
extrapolating the disc inwards from the relative exponential section
between radii of 50 to 130\arcsec. This disc is subtracted from the
total light to obtain the bulge surface brightness for radii less than
47\arcsec\ as has been illustrated in Fig. 12. Note that the bar
extends out to {72}\farcs{5} and hence in the bar region
the bulge slowly starts to build up coming closer to the centre
as suggested by the optical image. 
For the disc again the same sech-squared density distribution is
assumed while the bulge is taken to be spherical.
Purpose of the bulge/disc decomposition is to investigate
whether and how much fit parameters will change compared to
a pure disc fit. To that aim the decomposition need not be
very precise.
There is a difference between the bulge and disc concerning the conversion
to face-on brightnesses. The photometric profile of the disc along
the major axis is converted to face-on magnitudes by multiplying
with a cosine(inclination) factor. Since the bulge
is assumed to be spherical no conversion to face-on is needed.
At this stage, no
corrections for internal and galactic absorption have been
made and consequently all derived M/L ratios are
those as observed. The rotation of the stellar and gas disc is calculated
by the prescription of Casertano (1983) while the rotation of the
bulge is given by the equations of Kent (1986).

The decomposition of the observed rotation curve is performed
by fitting the sum of the rotation curves of the components to the
observed data in a least squares sense. In that way the best fit
is designated as the situation of minimum ${\chi}^2$ value. 
However, a least squares fit procedure assumes that the fitting
function is known a priori and the data points scatter in a Gaussian
way around that function. For rotation curves that is not valid.
Firstly a rotational functionality for the halo is adopted which need
not be correct. Secondly the procedure to determine the rotation
is an approximation in the sense that azimuthal symmetry is assumed
with no in or outflow. For example spiral arms can produce small 
irregularities, which may lead to small systematic deviations from
the actual rotation law. Because of these matters the resulting
minimum ${\chi}^2$ value is only a limited indicator of the quality
of the fit. In general one has to make an inspection by eye to judge
the quality of the fit, taking
the errors of the individual data points into account. 

First the decomposition results for a disc only situation will be considered.
Three parameters are free and have been fitted simultaneously: the core radius
and maximum rotation of the dark halo, and the M/L ratio of the disc.
The result is illustrated in Fig. 13a
and has a disc with an I-band mass-to-light ratio of 
1.79 $\pm$ 0.19. Further numerical values
are given now and for the following fits in Table 4. Surprisingly the
fit finds a least squares minimum for a non
maximum disc situation. This is surprising because in most cases
when rotation curves are decomposed a least squares fit 
tends to the maximum disc
solution while other solutions are nearly equally as good (van Albada
\& Sancisi 1986). The maximum rotation of the disc is 162 \kmss
which amounts to 60\% of the observed maximum rotation. This is close
to the 63\% found from the study of stellar velocity dispersions
by Bottema (1993). If the disc rotational contribution is forced below
50\% the agreement between data and model rotation rapidly becomes worse.
For such a situation features in the photometry cannot be reconciled
adequately with the accompanying features in the observed rotation curve.  
%
\begin{figure}
\resizebox{\hsize}{!}{\includegraphics{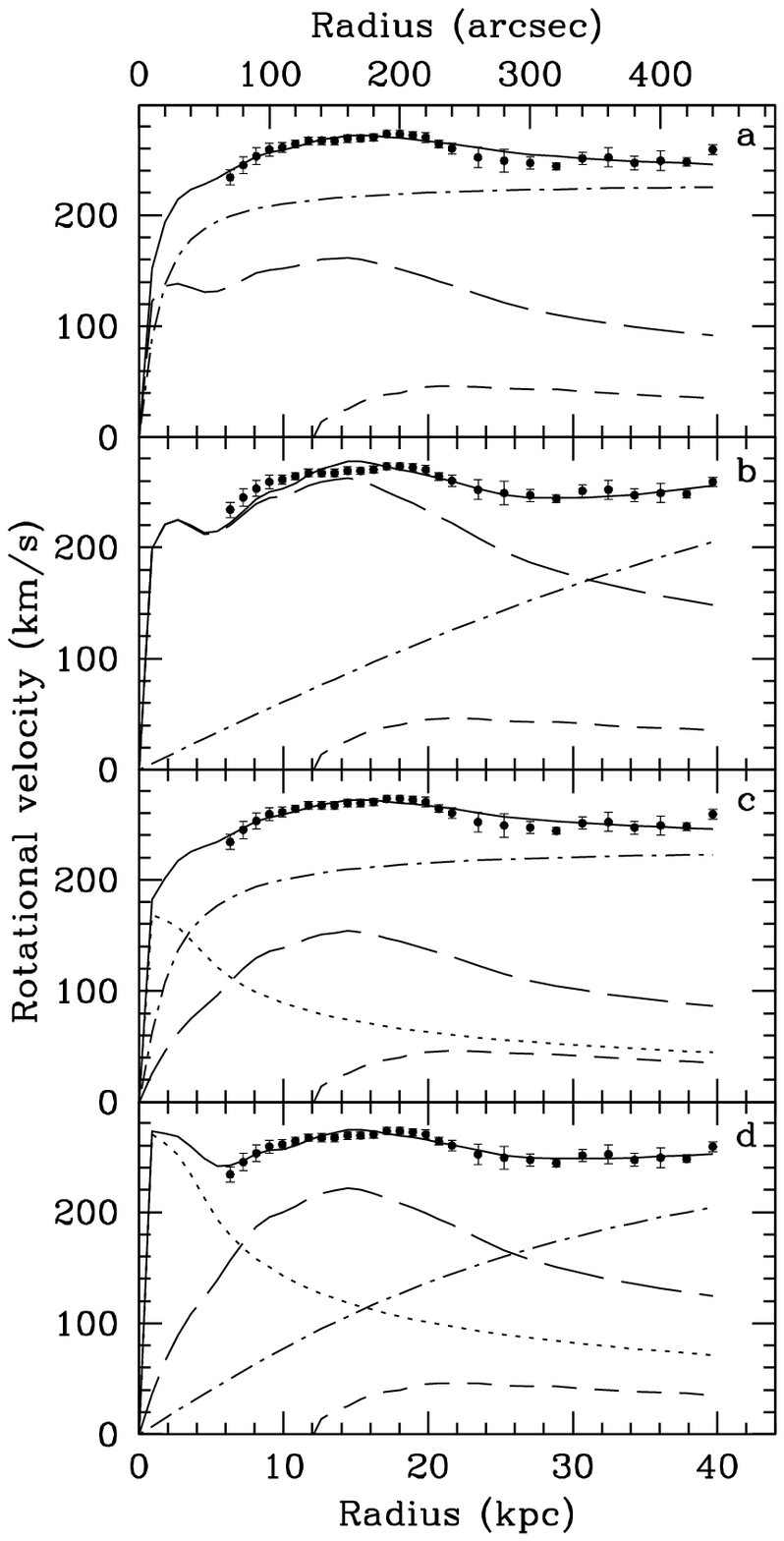}}
\caption[]{{\bf a to d.} Various rotation curve decompositions. The dots
are the observed rotational data. The fit
to these is indicated by the full drawn line.
Individual contributions of the bulge (dotted line), disc (long dashed line),
gas (short dashed line), and dark halo (dash - dot line)
are also given. Details are given in the text and in Table~4.
{\bf a:} 
All the mass is assumed to be in
a disc like distribution. The best fit is for a
disc contributing $\sim$ 60\% at most to the total rotation.
{\bf b:}
As {\bf a}, but now for a secondary minimum in the
least squares fitting procedure. This is a maximum disc fit.
{\bf c:}
For a separate bulge and disc mass distribution, where
the M/L ratios of both are constrained to be equal.
{\bf d:}
As {\bf c}, but the M/L ratios of bulge and disc are both unconstrained.
}
\end{figure}

The photometric radial profile clearly indicates that NGC 3992
is highly non exponential, contrary to claims by Elmegreen \& Elmegreen
(1985). This galaxy is of an extreme Freeman type II (Freeman 1970)
and the maximum of the rotation does not occur at the usual 2.2
scalelengths, if a scalelength can be defined at all.
Instead the relative exponential part between 50 and 130\arcsec\
followed by the more steep decline in brightness beyond, results
in a specific shape of the rotation curve of the luminous
matter as shown by the long dashed line in Fig. 13. This
curve has a maximum around 13 kpc and a long linearly decreasing
section between 17 and 24 kpc.
The observed drop in
the rotation curve is then not a consequence of a truncation feature
of the disc as suggested by G84 but simply a consequence of the
non-exponentiality of the disc. Also the photometry does not suggest
a sudden drop in brightness at a radius of $\sim$ 19 kpc = 211\arcsec\
needed to explain such a drop at the observed radii between
19 and 24 kpc. Some experimenting has been done by taking cut offs in
the photometry at 250\arcsec\ = 22.5 kpc and at larger radii.
But this is already so far out that an associated drop
in the rotation curve cannot be noticed any more (Casertano 1983).

There is a secondary minimum in the least squares fit, which is
illustrated in Fig. 13b. It is in principle a maximum disc fit
with a dark halo having a core radius comparable to the maximum
radius to where the rotation curve is determined. Compared to the
previous fit, the outer data points seem to be better represented.
In the stellar disc regions however, the fit is worse and at certain
positions not compatible with the data. The reduced minimum ${\chi}^2$ value
for this fit is 1.94, for the previous fit it was 1.22. The mass-to-light
ratio in the I band, uncorrected for absorption amounts to 
4.71 $\pm$ 0.11 for this maximum disc fit.

When a bulge is added to the system there is one more free parameter,
namely the M/L ratio of the bulge. 
This parameter has been constrained in two ways;
a situation where the M/L ratio is taken equal to that of the disc
(see Fig. 13c)
and a situation where the maximum rotation of the bulge is fixed
at the flat level of the rotation curve.
The fit is performed and dark halo parameters and disc M/L ratios
follow and are given in Table 4. 
The reduced ${\chi}^2$ value for both bulge M/L ratios
are equal. What one can achieve by adding a bulge is some mass transfer
from the dark halo to the bulge resulting in a larger core radius
compared to the disc only case. Presently core radii are 1.8 and
3.7 kpc compared to the 1.16 kpc when only a disc is present.
When the M/L ratio of the bulge is not constrained the fitting 
procedure generates a result given in Fig. 13d which is analogous
to the maximum disc fit in Fig. 13b. This fit is actually better then
the one where the M/L ratio of the bulge is constrained in the sense
that the reduced ${\chi}^2$ value is slightly lower. 
However the M/L ratios are substantial; 5.1 and 4.2
for the bulge and
disc respectively. So adding a bulge has
the side effect that one is again close to the situation for other
rotation curve decompositions: nearly equally good fits can be
made for a whole range of disc contributions to the total rotation.
Other indicators are then
needed to determine the M/L ratio of the disc, like observations of
disc stellar velocity dispersions or population synthesis arguments.

One can conclude that the features in the rotation curve are
generated by the strong non-exponentiality of the photometry.
Still, even with the determined detailed rotation 
curve of NGC 3992 it is not possible to get a tight constraint
on the disc contribution to the total rotation or on 
the M/L ratios of disc and bulge.
There is only a slight preference for the disc to be 
sub maximal.
\section{The mass-to-light ratio}
To put the mass-to-light ratio of NGC 3992 into perspective one should
compare it with values of other galaxies. Palunas \& Williams (2000)
give maximum disc $M/L$ I-band ratios for a sample of 74 spiral 
galaxies. The luminosities of their galaxies have been
corrected for absorption as if observed face-on. Unfortunately this
correction does not take into account the demonstrated dependence
of such a correction on total luminosity of a galaxy 
(Giovanelli et al. 1997; Tully et al. 1998).
In practice Palunas \& Williams make a correction more or less
as if all their galaxies belong to the most luminous category.
To circumvent this problem, 32 galaxies of the sample were selected
having a $^{10}$log$(2v_0)$ $>$ 2.6 or a maximum rotation 
approximately larger than 200 \kms. For that subsample the
average $(M/L)_I^{i-0}$ (so corrected to face-on) is 3.1 $\pm$ 0.7
with a total range of 2.2 to 4.3. There is another problem, however.
The sample is concentrated in the great attractor region and
galaxies or groups of galaxies may have large peculiar velocities
rendering the adopted Hubble flow distances uncertain. For example,
galaxies in the vicinity of the Centaurus cluster have a median
$(M/L)_I^{i-0}$ of 1.8 $\pm$ 0.6 while galaxies near the Hydra cluster
have a median $(M/L)_I^{i-0}$ of 3.4 $\pm$ 1.0.

As mentioned in the introduction, the $\kprime$-band rotation curve
decomposition of NGC 3992 by Verheijen (1997) indicates that
this galaxy has an abnormally large $(M/L)_{\kprime}$ value compared
to the other HSB galaxies of the UMa cluster. Is this also 
the case for the present more precise rotation curve and the 
I-band photometry? In Paper II (Bottema 2002)
a conversion is made to absorption corrected
face-on $(M/L)_I^{i-0}$ maximum disc values for ten UMa cluster
HSB galaxies not including NGC 3992. The average value of these ten
galaxies, $\langle (M/L)_I^{i-0} \rangle = 1.65$ with a range of
0.7 to 2.2. The same absorption procedure for NGC 3992 requires that
the observed $(M/L)_I$ ratio has to be divided by a factor 1.27. Then
the maximum disc fit (Fig. 13b) has a $(M/L)_I^{i-0}$ of 3.68 $\pm$ 0.1.

Compared to the other HSB galaxies of the UMa cluster the 
mass-to-light ratio is very large, also in the I-band. However, 
when compared with $(M/L)_I^{i-0}$ = 3.1 $\pm$ 0.7 for the 
luminous galaxies of the Palunas \& Williams sample the value of
NGC 3992 is still large, but not exceptional. When the notion holds
that barred galaxies are closer to maximum disc, then one would
expect (on average) that maximum disc $M/L$ ratios of barred galaxies
are determined to be 
smaller than those of non-barred galaxies. Consequently NGC 3992
contradicts this notion. Remains to be explained why the 
mass-to-light ratios of the different clusters can be so different. 

A comparison with $M/L$ ratios generated by population synthesis
models is in principle not possible. Those ratios depend on the
employed IMF, and in particular on the low mass end of the IMF.
Changing the functionality or the low mass cutoff also changes
the $M/L$ ratio (Jablonka \& Arimoto 1992; Bottema 1997;
Bell \& de Jong 2001). A detailed discussion of the relation
of NGC 3992 to its companions and its position on the
Tully-Fisher relation is deferred to Paper II.

\section{Discussion, outlook, and conclusions}

The present observations demonstrate that the amount of gas
in the bar region is very small. It is likely that any available
gas has been transported inwards as a result of shocks at the leading
side of the bar (Athanassoula 1992). Since the gas density is so low
this implies that the bar must be rather long lived in order to
have had enough time to transport all the gas to the centre. In addition
this means that the galaxy, at least the central region, must be
undisturbed which is substantiated by the very regular optical image
of the galaxy.

Even though NGC 3992 exhibits some specific features in its rotation
curve, a decomposition cannot give a tight constraint on the
contribution of the disc to the total rotation. As noted above,
the determined maximum disc M/L ratio is rather large compared to 
other galaxies while one might have expected the opposite. For such
a problem there is always a way out: put NGC 3992 at a larger distance
and so behind the UMa cluster. To make the M/L ratio 
equal to that of
the highest values of the other but similar galaxies of the cluster,
the distance to NGC 3992 has to be increased from 18.6 to 
$\sim$ 24 Mpc. To make the M/L ratio equal to the average
value of the other HSB galaxies, the distance needs to be $\sim$
28 Mpc. That distance would still be compatible with the
position of NGC 3992 and the position of its companions on the
Tully-Fisher relation (see Paper II).

In a recent numerical study of normal galaxies, including
gas and star formation (Bottema, in preparation) it is
demonstrated that for discs with an average thickness the bar
instability sets in for $v^{\rm max}_{\rm disc}/v^{\rm max}_{\rm obs}
\ga 0.8$. On the other hand for a relatively thin disc, which
implies a disc with a low stellar velocity dispersion, the bar
instability already occurs at lower 
$v^{\rm max}_{\rm disc}/v^{\rm max}_{\rm obs}$ values.
Therefore, if NGC 3992 would have a less massive disc the 
formation of the galaxy must have been rather specific.
Although we now enter the realm of speculation, one might
for example imagine a slowly and gently forming disc with modest gas content.
Star formation is low and disc heating by the few molecular clouds
proceeds slowly. In such a cold stellar disc, even if it is not maximal
a bar can be generated and in the absence of substantial gas accretion
can remain for a time comparable to the lifetime of the galaxy.

To further investigate the matter of bar existence and
formation one should resort to numerical calculations.
As mentioned above the main parameters governing the bar
instability of a disc are the ratio of dark to luminous matter
and the disc thickness. Unfortunately matters are more complicated
because the past history of a galaxy will be of influence
on its present morphology.

Not only for NGC 3992 but also for galaxies in general
the main question which remains unanswered is: What is the precise
ratio of dark to luminous matter? With the advent of large telescopes
it now becomes feasible to do more detailed and more extensive
observations to determine the stellar velocity dispersions
of galactic discs. In that way the results for the
sample of Bottema (1993) should be checked and extended.

Because of the large amount of observational material
it was decided to split up the description of the NGC 3992 group
in two parts. This part (Paper~I) deals with the main galaxy
and focuses on its barred nature and mass distribution.
In Paper~II the observations of the three small companions
are described. Velocity fields and rotation curves are derived
and rotation curve decompositions have been made.
For all the four galaxies of the group an analysis is presented
of colours, M/L ratios, and position of the galaxies in 
the TF relation of the Ursa Major cluster. 

Finally a compilation of the main conclusions of this paper:

\begin{description}
\item[{\bf 1.}]
Detailed observations in the neutral hydrogen line have
been made of the large barred spiral galaxy NGC 3992
and its three small companion galaxies, UGC 6923, 
UGC 6940, and UGC 6969.
\item[{\bf 2.}]
In general the \hi distribution of NGC 3992 is regular;
it has a faint radial \hi extension outside its stellar disc.
\item[{\bf 3.}]
There is a pronounced central \hi hole in the gas distribution
at exactly the radial extent of the bar.
\item[{\bf 4.}]
It is likely that any available gas has been transported inwards
by the bar. Because of the emptyness of the hole no major gas 
accretion events can have occurred in a recent galactic period.
\item[{\bf 5.}]
The distortions generated by the bar on the velocity field
are limited to its proximity and are only minor. 
\item[{\bf 6.}]
From the velocity field a detailed and extended 
rotation curve has been derived
which shows some distinct features.
\item[{\bf 7.}]
These distinct features can be explained by
the non-exponential radial light distribution of NGC 3992.
\item[{\bf 8.}]
A rotation curve decomposition gives a slight preference
for a sub maximal disc, though a range of disc contributions
until a maximum disc situation can give a nearly equally good representation
of the rotation curve.  
\item[{\bf 9.}]
In case of such a maximum disc the mass-to-light ratio is large
but not exceptional. 
\end{description}

\begin{acknowledgements}
The observations presented in this paper were obtained with
the Westerbork Synthesis Radio Telescope (WSRT) which is operated
by the Netherlands Foundation for Research in Astronomy (NFRA).
We thank R. Sancisi, J. Gerssen, and I. Garcia-Ruiz
for insightful discussions.
R.B. thanks the Kapteyn Institute for hospitality and support.
\end{acknowledgements}

\clearpage

\setcounter{table}{3}
\begin{table*}
\caption[]{Decomposition of the rotation curve of NGC 3992}
\begin{flushleft}
\begin{tabular}{lllllllll}
\noalign{\smallskip}
\hline
\noalign{\smallskip}
Situation & panel in & red & disc & (M/L)$_{\rm d}$ & bulge &
 (M/L)$_{\rm b}$ & $\rcore$ & $v_{\rm h}^{\rm max}$ \\
  & Fig. 13 & ${\chi}^2$ & mass & & mass & & & \\
  & & & (10$^9 M_{\sun}$) & ($M_{\sun}/L_{\sun}^I$) & (10$^9 M_{\sun}$) &
  ($M_{\sun}/L_{\sun}^I$) & (kpc) & (\kms) \\
\noalign{\smallskip}\hline\noalign{\smallskip}
D only, best fit & a & 1.22 & 73.7 & 1.79 $\pm$ 0.19 & - & - & 
  1.16 $\pm$ 0.35 & 230 $\pm$ 98 \\
D only, max disc & b & 1.94 & 194.1  & 4.71 $\pm$ 0.11 & - & - & 
  44.9 $\pm$ 17  & 482 $\pm$ 188 \\
D + B, equal M/L & c & 1.22 & 64.9 & 2.03 $\pm$ 0.21 & 18.7 & 2.03 & 
  1.79 $\pm$ 0.35 & 230 $\pm$ 64 \\
D + B, $v^{\rm max}_{\rm b}$ = 240 & - & 1.25 & 71.3 & 2.23 $\pm$ 0.26 & 
  36.9 & 4.0 & 3.7 $\pm$ 0.6 & 233 $\pm$ 57 \\
D + B, max & d & 1.08 & 134.6 & 4.2 $\pm$ 0.3 & 47.1 & 5.1 $\pm$ 0.5 & 
  23.2 $\pm$ 5.7 &  327 $\pm$ 91 \\
\noalign{\smallskip}\hline
\end{tabular}
\end{flushleft}
\end{table*}

\setcounter{table}{1}
\begin{table}
\caption[]{Observing parameters}
\begin{flushleft}
\begin{tabular}{ll}
\noalign{\smallskip}
\hline
\noalign{\smallskip}
Telescope & WSRT \\
Observing date & May 1997 to Sept. 1997 \\
Duration of observation & 4 $\times$ 12 h. \\
Number of interferometers & $\sim$ 27 \\
Baselines (min-max-incr.) & 36 - 2736 - 36 m. \\
Full res. beam (FWHM, $\alpha \times \delta$) & 14\arcsec $\times$ 18\arcsec \\
FWHpower primary beam & 37\arcmin \\
Rms (1$\sigma$) noise per channel & \\ 
\quad full res. & 1.96 K = 0.473 $M_{\sun}$pc$^{-2}$ \\
\quad res. = 30\arcsec $\times$ 30\arcsec & 0.55 K = 0.132 $M_{\sun}$pc$^{-2}$ \\
Velocity central channel & 1050 \kms \\
Bandwidth & 5 MHz \\
Number of channels & 64 \\
Channel separation & 16.6 \kms \\
Velocity resolution & 33.3 \kms \\
Field centre (1950) & (11$^{\rm h}$ 55$^{\rm m}$ 07$^{\rm s}$ ; 53\degr 39\arcmin 18\arcsec) \\
K-mJy conversion, & \\
\quad equivalent of 1 mJy/beam & 2.62 K (full res.) \\
 & 0.73 K (res. = 30\arcsec) \\
Adopted distance & 18.6 Mpc \\
\noalign{\smallskip}
\hline
\end{tabular}
\end{flushleft}
\end{table}

\setcounter{table}{2}
\begin{table}
\caption[]{The rotation curve of NGC 3992}
\begin{flushleft}
\begin{tabular}{llllll}
\noalign{\smallskip}\hline\noalign{\smallskip}
R & $V_{\rm rot}$ & ${\varepsilon}_{\rm vrot}$ &
R & $V_{\rm rot}$ & ${\varepsilon}_{\rm vrot}$ \\
(\arcsec) & (\kms) & (\kms) & (\arcsec) & (\kms) & (\kms) \\
\noalign{\smallskip}\hline\noalign{\smallskip}
70 & 234 & 6.8 & 210 & 272 & 3.1 \\
80 & 245 & 7.8 & 220 & 270 & 4.3 \\
90 & 253 & 7.3 & 230 & 264 & 3.4 \\
100 & 259 & 6.0 & 240 & 260 & 5.0 \\
110 & 261 & 4.3 & 260 & 252 & 9.0 \\
120 & 264 & 3.2 & 280 & 249 & 10.4 \\
130 & 267 & 3.3 & 300 & 247 & 5.5 \\
140 & 267 & 3.5 & 320 & 244 & 3.2 \\
150 & 267 & 3.4 & 340 & 251 & 5.5 \\
160 & 269 & 3.4 & 360 & 252 & 8.6 \\
170 & 269 & 3.2 & 380 & 247 & 5.9 \\
180 & 270 & 3.1 & 400 & 249 & 8.7 \\
190 & 273 & 3.1 & 420 & 248 & 3.5 \\
200 & 273 & 3.1 & 440 & 259 & 4.3 \\
\noalign{\smallskip}\hline\noalign{\smallskip}
\multicolumn{6}{l}{Pos. of dynamical centre}\\
\multicolumn{3}{l}{\quad RA (1950)}&\multicolumn{3}{l}{
11$^{\rm h}$ 55$^{\rm m}$ {0}\fs{59} }\\
\multicolumn{3}{l}{\quad Declination (1950)}&\multicolumn{3}{l}{
53\degr 39\arcmin {10}\farcs{9}  }\\
\multicolumn{3}{l}{\quad $V_{\rm sys}$ (Hel.)}&\multicolumn{3}{l}{
1049 $\pm$ 2 \kms}\\
\multicolumn{3}{l}{Inclination}&\multicolumn{3}{l}{57\degr $\pm$ 1\degr}\\
\multicolumn{3}{l}{P.A.}&\multicolumn{3}{l}{
245\degr $<$ P.A. $<$ 255\degr}\\
\noalign{\smallskip}\hline
\end{tabular}
\end{flushleft}
\end{table}

\setcounter{table}{0}
\begin{table}
\caption[]{Galaxy parameters}
\begin{flushleft}
\begin{tabular}{lll}
\noalign{\smallskip}
\hline
\noalign{\smallskip}
\quad NGC 3992 & & \\
Hubble type & SBb(rs)I & a \\
Brightness (in B) & 10.86 mag. & b \\
Brightness (in I) & 8.94 mag. & b \\
Opt. incl. ($q_0 = 0.11$) & 57\degr & b \\
Opt. PA major axis & 68\degr (= 248\degr) & b \\
PA major axis bar & 37\degr & c \\
Deprojected bar length & 145\arcsec & d \\
Scalelength & undef & \\
Total \hi mass & 5.9 10$^9$ \msol & d \\
21 cm cont. flux & 43.2 mJy & d \\
\noalign{\medskip}
\quad UGC 6923 & & \\
Brightness (in B) & 13.91 mag. & b \\
Brightness (in I) & 12.36 mag. & b \\
Opt. incl. ($q_0 = 0.11$) & 66\degr & b \\
Opt. PA major axis & 354\degr & b \\
Scalelength (in I) & {20}\farcs{9} & b \\
Total \hi mass & 0.64 10$^9$ \msol & e \\
21 cm cont. flux & $<$ 2.6 mJy & b \\
\noalign{\medskip}
\quad UGC 6940 & & \\
Brightness (in B) & 16.45 mag. & b \\
Brightness (in I) & 15.44 mag. & b \\
Opt. incl. ($q_0 = 0.11$) & 75\degr & b \\
Opt. PA major axis & 135\degr & b \\
Scalelength (in I) & {8}\farcs{52} & b \\
Total \hi mass & 0.16 10$^9$ \msol & e \\
21 cm cont. flux & $<$ 1.3 mJy & b \\
\noalign{\medskip}
\quad UGC 6969 & & \\
Brightness (in B) & 15.12 mag. & b \\
Brightness (in I) & 14.04 mag. & b \\
Opt. incl. ($q_0 = 0.11$) & 73\degr & b \\
Opt. PA major axis & 330\degr & b \\
Scalelength (in I) & {11}\farcs{65} & b \\
Total \hi mass & 0.44 10$^9$ \msol & e \\
21 cm cont. flux & $<$ 3.8 mJy & b \\
\noalign{\smallskip}
\hline
\multicolumn{3}{l}{a Sandage \& Tammann (1981)} \\
\multicolumn{3}{l}{b Verheijen (1997)} \\
\multicolumn{3}{l}{c Measured from photograph} \\
\multicolumn{3}{l}{d This paper} \\
\multicolumn{3}{l}{e Paper II} \\
\end{tabular}
\end{flushleft}
\end{table}

\end{document}